# Build-at-home UV-C disinfection system for healthcare settings

Rosemary C. She[1], Dongyu Chen[2], Pil Pak[1], Deniz K. Armani[3]*, Andreas Schubert[3], Andrea M. Armani[2,4]*

[1] Department of Pathology, Keck School of Medicine of the University of Southern California, Los Angeles, CA 90033

[2] Ming Hsieh Department of Electrical Engineering, University of Southern California, Los Angeles, CA 90089

[3] SMP Engineering, 1805 Flower Street, Glendale, CA 91201

[4] Mork Family Department of Chemical Engineering and Materials Science, University of Southern California, Los Angeles, CA 90089

*darmani@smg-eng.com; armani@usc.edu

## Abstract

Significant research has shown that UV-C exposure is an effective disinfectant for a range of bacteria and viruses, including coronaviruses. As such, a UV-C treatment in combination with a chemical wipe, such as EPA hydrogen peroxide, is a common cleaning protocol in a medical setting, and such disinfection protocols have gained in importance during the current COVID-19 pandemic due to the need to reuse PPE. However, given the substantial increase in patient volume, the quantity of materials requiring disinfection exceeds the UV-C equipment throughput capabilities at many medical facilities. Therefore, there is a need for a UV-C disinfection system that can be rapidly deployed. In response to this demand, we designed, constructed, and validated a UV-C disinfection system from readily accessible components; specifically, a plastic bin, UV-C light bulb and conventional light housing. To further improve the performance, the interior of the tub was spray-painted with chrome paint, forming a low quality-factor (Q) fabry-perot optical cavity. As part of this work, a set of modular design criteria which allows for flexibility in component selection without degradation of UV-C dose performance is established. This flexibility is critical given the current fluctuating availability of source materials. The disinfection capabilities of the system are validated using *Bacillus cereus*, a gram-positive endospore-forming bacteria.

## Introduction

Highly infectious microbial and viral diseases are a major challenge to global health, and, as such, are also a significant risk to global financial stability and security. While vaccines play a key role in preventing viral epidemics and pandemics, once an outbreak has occurred, the implementation of disinfection measures to limit spread becomes paramount. As a result, numerous methods of disinfection have been developed, including chemical (e.g. EPA hydrogen peroxide), radiation (ozone, UV-C, microwave), and thermal.[1–6] Among these different types,

UV-C disinfection has gained favor due to its efficacy against a broad range of microbial and viral agents in a variety of environments.[3,7,8]

The UV-C wavelength band covers 100nm-280nm, and it directly overlaps with the peak absorption of DNA and RNA (~260nm). The inactivation mechanism is straight-forward. Upon UV-C absorption, the pyrimidines in the RNA or DNA are converted to pyrimidine (6–4) pyrimidone photoproducts and cyclobutane pyrimidine dimers.[9] If the population of dimers is sufficiently high, transcription errors occur, ultimately resulting in inactivation of the bacteria or virus. As such, UV-C is a nearly universal disinfection method for bacteria, and its effectiveness in viral disinfection is not correlated with virus size, but with pyrimidine concentration.

Previous works have shown 3 log inactivation using UV-C of a range of viruses including MMV, PPV, PRV, BVDV, and H1N1[7–10]. However, this previous work relied on large commercial systems that can be challenging to procure, particularly in the midst of a pandemic or in a low resource environment. Thus, given the universality of UV-C as a disinfection method, designing and validating a system that can be constructed from easy to procure components has significant societal impact.

In this work, we demonstrate a home-built UV-C disinfection system based on commonly available components, including a plastic bin, UV-C light bulb, and standard light housing (Figure 1). To improve the intensity of the UV-C light inside the box, a low-Q Fabry-Perot cavity is formed by coating the box interior with a reflective coating using chrome spray paint. The efficacy is validated using *Bacillus cereus* as a test organism. *B. cereus* is an aerobic, rod-shaped, gram-positive bacteria that can quickly multiply at room temperature. It can cause several illnesses primarily due to its production of several tissue-destructive exoenzymes. *B. cereus* can form endospores that can withstand harsh conditions including UV exposure.[11,12]

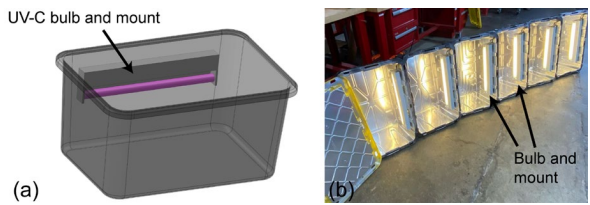

Figure 1: UV-C Disinfection system. (a) Schematic of system. (b) Example of several systems on their sides. Shown with visible lights for safety.

## Theory

The most important element of the system is optical power delivered by the UV-C source. As simple guidelines, the dose delivered by the UV-C source is dependent on the distance the object is away from the source (l), the source wattage and efficiency, and the exposure duration. In other words, a lower wattage source can be used in place of a higher wattage source, if the exposure duration is increased. Previous work has shown that UV-C doses in the 10mJ/cm$^2$ are capable of a 3 log reduction in bacteria growth[8]. In a virus sample, an order of magnitude higher dose is typically required to achieve the same reduction.

Given that access to components may be highly variable, it is desirable to create a generalizable expression governing the UV-C intensity and intensity distribution inside of the UV-C system. This expression can then be used to calculate the impact of different bulbs on UV-C dose, accelerating re-design when needed.

The UV-C bulb is treated like a linear light source, where each point on the bulb is a point light source with isotropic radiation. The cumulative UV-C dose created by the bulb is an integration of the dose generated by all the point light sources on the bulb.

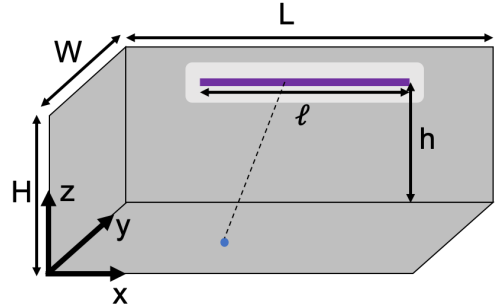

Figure 2: Schematic showing key variables used to calculate cumulative UV-C dose.

As is shown in Figure 2, assuming we have an enclosure with length L, width W and height H. The bulb is placed on the side wall above the floor of the box by a height of h, and it has an optically active length of $\ell$ and is centered on the wall horizontally. Given this configuration, for any point source on the bulb, the location of the point source is $(X, W, h)$, where $X \in [\frac{L-l}{2}, \frac{L+l}{2}]$. For any point $(x, y, z)$ in the box, the distance between the point light source and the detection point is:

$$Distance = \sqrt{(X-x)^2 + (W-y)^2 + (h-z)^2}$$

Assuming the UV-C wattage power of the bulb is $P$ (unit: Watt), then the power of the point source is $P/\ell$. As the point source radiates isotropically in a sphere, the radiation intensity on point $(x, y, z)$ would be:

$$\frac{P/\ell}{4\pi((X-x)^2 + (W-y)^2 + (h-z)^2)}$$

By integrating all the point sources on the bulb, we get the intensity generated by the whole bulb:

$$I = \int_{\frac{L-l}{2}}^{\frac{L+l}{2}} \frac{P/\ell * dX}{4\pi((X-x)^2 + (W-y)^2 + (h-z)^2)}$$

The intensity calculated above is for the situation where there is no reflection from the box. However, in the system developed in the present work, the interior is coated with a reflective chrome paint. To simplify the reflection calculation, we only take the first order of reflection into consideration. In other words, the reflection of the reflected light is not considered. The initial calculation of intensity generated by the reflection is similar to the previous one with a slight modification. To account for the reflection, a series of "virtual light bulbs" with a power of $\alpha P$, where $\alpha$ is the reflection rate, are located on the walls counter to the primary light source. The intensity generated by each mirrored bulb can also be calculated using the integration method. The only difference is that the distance between point $(x, y, z)$ and the mirrored bulb is different from the previous calculation. By adding the reflection caused by all the walls (including ceiling and floor) together, we get the intensity of the UV-C light inside the box. The intensity can be expressed in the following equation:

$$I_{with\ reflection} = \int_{\frac{L-l}{2}}^{\frac{L+l}{2}} \frac{P/l * dX}{4\pi \left((X-x)^2 + (W-y)^2 + (h-z)^2\right)} + \sum_{reflective\ walls} \int_{\frac{L-l}{2}}^{\frac{L+l}{2}} \frac{\alpha P/l * dX}{New\ distance}$$

In the system shown in Figure 1, L = 78.74 cm, W = 50.8 cm, H = 35.56 cm, h = 25.4 cm, and $l$ = 43.18 cm. The wattage of the UV-C light bulb is 15 W, and, considering a 35% conversion efficiency, the UV-C wattage $P$ = 5.25 W. By putting all these parameters into the equation, we can calculate the intensity at any point in the box (unit: W/cm$^2$). To determine the UV-C dose, this value should be multiplied by the time that the system is on.

Figure 3 shows a series of calculations for an exposure time of 3 minutes. The wall reflectivity is varied between 0% and 50%. While the system is able to exceed the 10mJ/cm$^2$ target dose with the 3 minute exposure throughout the box in all three systems, with a 50% reflectivity, the dose delivered is several orders of magnitude higher. This increase would allow for a substantial decrease in exposure duration, accelerating throughput of the system.

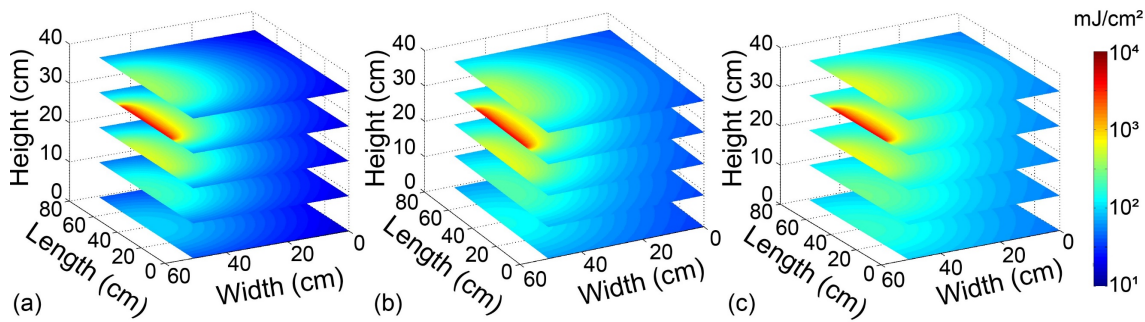

Figure 3: Cumulative UV-C dose delivered for a three minute exposure inside the UV-C disinfection system. Three different wall reflectivity values are modeled: (a) 0%, (b) 25%, and 50%.

**Results**

As can be seen in Figure 4, the *B. cereus* samples that were not exposed to UV-C readily formed colonies after 24 hour incubation time indicating that all preparation methods were correct. The baseline counts of Petri dish organism films ranged from 1 to 3 x 10$^6$ colony forming units (CFU). In contrast, this growth was dramatically eliminated when the samples were exposed to the UV-C, even with only 1 minute exposure (Figure 4, inset).

At 1 minute UV-C exposure, all organism counts were nil except for one of three replicates in which 800 CFU remained. At 3 and 6 minute UV-C exposure times, organism counts were reduced to undetectable levels except for one of three replicates in which 100 CFU remained. Directly irradiated organisms on blood agar plates demonstrated no growth of organisms at 3 and 6 minute exposure times but growth of 700 CFU/mL after 1 minute exposure, whereas baseline count was >3 x 10$^5$ CFU/mL (Figure 4).

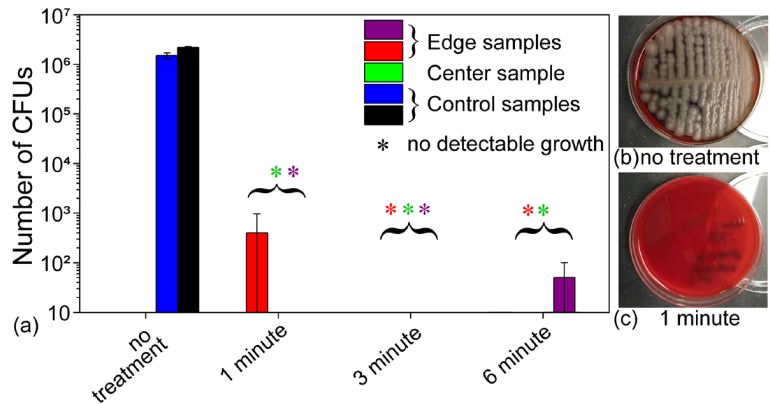

Figure 4: Summary of results. (a) In both control samples, over $10^6$ colony forming units (CFUs) grew during the 24 hour incubation period. In contrast, the majority of the exposed dishes were unable to support colony formation, indicating a 6 log reduction. (b) Control sample of *B. cereus* not exposed to UV-C formed colonies after 24 hour incubation time. (c) In contrast, this growth was dramatically eliminated after 1 minute exposure of identical sample preparations to the UV-C.

In addition to achieving >3 log reduction, the exposure times compare favorably to commercial UV-C disinfection systems which typically have disinfection protocols between 1-2 minutes. The alignment between the current build-at-home approach and the conventional commercial systems is possible because the cumulative dose delivered is the same, as shown in Figure 3. However, the proposed system offers numerous advantages. Specifically, the design can be adapted to use any UV-C bulb length, it is portable and significantly lighter, and it is less expensive.

## Conclusions

In conclusion, we have designed and validated a simple to construct UV-C disinfection system. It uses readily available, inexpensive components to create the UV-C chamber. The efficiency of the system is further improved by turning the chamber into a low-Q Fabry-Perot cavity by coating the interior with a reflective material. As a result, the intensity of the UV-C optical field is amplified, allowing shorter exposure times to be used. Over 3 log reduction in CFUs is confirmed using *B. cereus* as a model bacteria. This approach will find use during the current COVID-19 pandemic where PPE is in short supply as well as in the future in low resource environments.

## Methods

Before presenting the methods, the authors would like to emphasize that exposure to UV-C has been directly and indirectly linked to several diseases and life-threatening conditions. Therefore, it is critically important to avoid indirect or direct exposure to the UV-C source.

Box fabrication: The interior of a 17 gallon plastic tote (Home Depot) was first cleaned with Acetone and then coated with chrome spray-paint (Rustoleum, Bright coat Metallic Chrome, Home Depot). The UV-C source was an 18" Philips UV-C bulb (G-15T8, 15W, McMaster). Two holes were drilled in the side of the tote to mount an under-counter light fixture (Home Depot) which held the UV-C source. Power was supplied to the light fixture via a power cord which was run through a third hole, located in the corner. To control the light externally, the power cord was plugged into a power strip with an on/off switch.

Bacteria growth/sample preparation protocol: A *B. cereus* isolate previously isolated from routine clinical culture was selected for this study. After subculture on sheep blood agar and overnight incubation at 35°C, a 4.0 McFarland suspension was prepared in sterile 0.45% saline solution. For analysis of UV-C irradiation of plastic surface, 100 µL aliquots were dispensed onto sterile polystyrene Petri dishes (100 mm diameter) over an approximately 1 x 2 cm area, then allowed to completely dry at 35°C, up to 30 minutes. Three replicates were included for each exposure time. For analysis of direct irradiation of agar media, sheep blood agar plates were inoculated with 1 µL and 10 µL of the 4.0 McF suspension prior to UV-C exposure. Baseline counts were obtained for each experiment and prepared in the same manner as experimental samples.

Directly after UV-C exposure of organism films on Petri dishes, each dish was flooded with 10 mL of 0.45% saline and scraped to completely resuspend the film. For baseline counts, 100 µL of 1:100 and 1:10,000 dilutions were plated in duplicate. For UV-C irradiated organism counts, 100 µL of neat, 1:100, and 1:10,000 dilutions of each sample were plated in duplicate. Organism counts were performed after 24 hr incubation of culture plates at 35°C in 5% $CO_2$.

## Acknowledgements

The authors would like to thank Dr. Kymberly Lengyel and Dr. Neha Nanda for their advice on system design.